\documentclass[useAMS,usenatbib]{mn2e}
\usepackage{graphicx}
\usepackage{amssymb}

\usepackage[dvipsnames]{color}

\def\ltsima{$\; \buildrel < \over \sim \;$}
\def\simlt{\lower.5ex\hbox{\ltsima}}
\def\gtsima{$\; \buildrel > \over \sim \;$}
\def\simgt{\lower.5ex\hbox{\gtsima}}

\title[Discs to DM haloes interaction]{Interaction between collisionless galactic discs and nonaxissymmetric dark matter haloes}

\author[Khoperskov et al]{A.V.~Khoperskov,$^1$ S.A.~Khoperskov,$^{2,3}$ A.V.~Zasov,$^3$  D.V.~Bizyaev,$^{3,4}$ S.S.~Khrapov$^1$\\
  $^1$Volgograd State University, Universitetsky pr., 100, 400062, Volgograd, Russia\\
  $^2$Institute of Astronomy Russian Academy of Sciences, Pyatnitskaya st., 48, 119017, Moscow, Russia \\
  $^3$Sternberg Astronomical Institute, Moscow M.V. Lomonosov State University, Universitetskij pr., 13, 119992, Moscow, Russia \\
  $^4$New Mexico State University and Apache Point Observatory, P.O. Box 59, Sunspot, NM 88349, USA}
\date{}

\pagerange{\pageref{firstpage}--\pageref{lastpage}} \pubyear{2012}

\def\LaTeX{L\kern-.36em\raise.3ex\hbox{a}\kern-.15em
    T\kern-.1667em\lower.7ex\hbox{E}\kern-.125emX}

\begin{document}

\label{firstpage}

\maketitle

\begin{abstract}
 Using $N$-body simulations ($N\sim 10^6 - 10^7$), we examine how a non-axisymmetric dark halo affects the dynamical evolution of the structure in collisionless (stellar)  discs. We demonstrate how the model parameters such as mass of the halo, initial conditions in the disc and the halo axes ratio affect  morphology and kinematics of the stellar discs. We show that a non-axisymmetric halo can generate a large-scale spiral density pattern in the embedded stellar disc. The pattern is observed in the disc for many periods of its revolution, even if the disc is gravitationally over-stable.  The growth of the spiral arms is not accompanied by significant dynamical heating of the disc, irrelevant to its initial parameters. We also investigate transformation of the dark halo's shape driven by the long-lived spiral pattern in the disc . We show that the analysis of the velocity field in the stellar disc  and in the spiral pattern gives us a possibility to figure out the spatial orientation of the triaxial-shaped dark halo and to measure the triaxiality.
\end{abstract}

\begin{keywords}
Galaxies: evolution; galaxies: haloes; galaxies: kinematics and dynamics; galaxies: spiral
\end{keywords}

\section{Introduction}

Both dynamic models of real galaxies and cosmological simulations give a strong evidence that the stellar discs of spiral galaxies are embedded into massive dark haloes. The halo mass and its influence on the disc evolution is the matter of debates. Among the different ways to put constraints on the  halo mass it is worth mention the
use of gravitational stability criteria in the disc's
plane~\citep{Safronov1960,Toomre-1964!Criterion-Toomre,Linden-Bell196,
Khoperskov-Zasov-Tiurina-2003!GravitInstab,
Zasov-Khoperskov-Tiurina-2004!Statist-dispers-vel}, a condition of stellar
disc  stability to bending perturbations
\citep{Zasov-Makarov-Michajlova-1991!N-body-vert,
Sotnikova-Rodionov-2004!Bending, Khoperskov-etal-2010!z-str}, simulation of polar rings
\citep{Whitmore-etal-1987!halo-polar-ring,SackettSparke1990,Reshetnikov-Sotnikova-2000!Polar-Ring-Halo,Iodice}, and the gravitational
lensing~\citep{Koopmans1998,Maller-etal-2000!Gravitational-Lensing,Oguri2003}.

Morphology of galactic spiral patterns also indicates
significant amount of the dark matter within the optical radius
\citep{Athanasoula-Bosma-Papaioannou-1987!Halo-spiral-gal-swing,
Korchagin-etal-2000!N1566, Griv-2004!Jeans-unstable}. Estimates of the dark mass contribution can also be obtained from the
vertical structure of stellar discs analysis (e.g.
\citet{Bahcall1984,Zasov-Makarov-Michajlova-1991!N-body-vert, BizyaevMitronova2002, ZasovBizyaev2002,
Korchagin-etal-2003!Local-Surface-Density,
BizyaevMitronova2009,Just2010,Garbari2011}). Maximum disc models suggest that $ M_h / (M_d +
M_b) \simeq 0.5 - 1 $, where $M_h$ is the halo mass within the optical radius, $M_b$ is the mass of stellar bulge, $M_d$ is the mass of stellar disc~\citep{Bottema1993}. Note that some
galactic discs (especially low luminous ones) are less massive than that required by the maximum
disc assumption~(\citet{Zasov-Khoperskov-Tiurina-2004!Statist-dispers-vel,
Bershady-etal-2011!Galaxy-Disks-Submaximal,
Westfall-etal-2011!Dark-matter-dominated-UGC463,
Saburova-2012!M33}).  At any case, one may conclude that in most galaxies (at least of normal surface brightness) masses $M_h$ and $M_d$ are comparable within the optical borders.

Modern cosmological experiments suggest a triaxial distribution of the dark matter in the haloes
(e.g. \citet{Zentner-2005!Anisotropic-distribution-galactic-satellites, Kuhlen-etal-2007!shapes-orientation-subhalo},
\citet{Muñoz-Cuartas2011}, \citet{Klypin2011}, and references therein).
Intriguing problem is a possibility to explain the generation and support for the long-lived and extended spiral structure in the stellar discs by triaxial
potential of the dark halo. True mechanism of the formation of the spiral structure is still a subject of discussions.
Although a bar or a close satellite may induce the spiral density wave, these events provide a relatively short term support.
In contrast, the dark halo can keep its triaxiality and support the spiral structure over much longer time scale.
The problem is to estimate whether the existing deviation from the
axial symmetry is {\it quantitatively} enough to support the spiral pattern in collisionless gravitationally stable discs.

Usually the distribution of the volume density in the halo is approximated as follows:
\begin{equation}\label{eq-shape-halo}
    \varrho_h(x,y,z) = \varrho_h(\xi) \,, \quad \xi\equiv
    \sqrt{\frac{x^2}{a^2} + \frac{y^2}{b^2} + \frac{z^2}{c^2}} \,,
\end{equation}
where the $ a=b=c $ case corresponds to the spherical symmetry in the density distribution. The model with $ a=b \ne c $ corresponds to an axisymmetric halo coplanar with the stellar disc.  Note that in this labeling convention, the relative
ordering of $a$, $b$, and $c$ is unconstrained.
The ratios $ q = b / a $, $ s = c / a $ describe the deviation of the density profiles front the spherical symmetry.
For simplicity, our further consideration of the dark matter (DM) haloes assumes
that $a \geq b$ and $c=a$, where $a$ and $b$ are the halo major and minor axes in the disc plane.

Cosmological $N$-body simulations show that  the triaxiality parameters $ s $ and $ q $ depend on the halo mass and distance to the centre~\citep{Bailin-Steinmetz-2005!Shapes-halo,Hayashi-etal-2007!shape-gravitational-potential-halo,Widrow-2008!Disk-Galaxies-Triaxial-Halos}.
Simulations conducted with a large number of particles  (as the Via Lactea with $N=(1-2.3)\cdot 10^8$,  \citet{Diemand-etal-2007!Dark-matter-substructure,  Kuhlen-etal-2007!shapes-orientation-subhalo,Debattista2008,DiemandMoore2009}) reveal that the dark haloes are more asymmetric at their central regions than at the periphery where the ratios $ b / a $ and $ c / a $ are close to  to unit.
For the topic of this paper the most important parameter of the halo's shape is the axes ratio $q$.

\begin{figure}\label{fig01}
  \includegraphics[width=0.9\hsize]{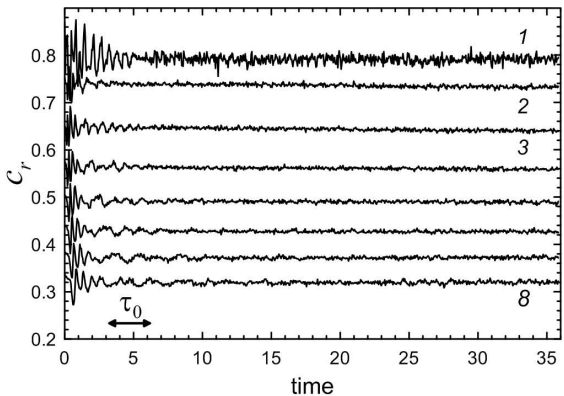}\caption{Evolution of the stellar velocity dispersion $ c_r $ at different radii $ r $ in a model with
  spherically symmetric halo. The curve \textit{1} demonstrates the central zone of the disc, the curve \textit{8} -- the periphery of the disc.
  The arrow indicates the time interval corresponding to one period of rotation $\tau_0$ at the disc's periphery.}
\end{figure}

Observing constraints on the dark halo geometrical parameters come from studies of globular clusters, variable stars (such as RR Lyrae), blue horizontal branch stars, low-metallicity giants, red giants, Cepheids (W Vir type) and F-dwarfs~\citep{Keller-etal-2008!Substructure-Galactic-Halo}.  Anisotropy of the velocity dispersion of
objects in the stellar halo of our Galaxy indicates an asymmetric distribution of DM~\citep{Morrison-etal-2009!MW-Inner-Halo}. Stellar haloes in other spiral galaxies also demonstrate significant deviation from spherical symmetry in the haloes: $ c / a \sim 0.3$ is suggested by \citet{Cooper-etal-2009!Galactic-stellar-haloes-CDM-model}. According to the Sloan Digital Sky Survey (SDSS) and Sloan Extension for Galactic Understanding and Exploration results, our galactic spheroid has a triaxial shape
with $ s = 0.63 $, $q=0.75$~\citep{Newberg-etal-2007!Asymmetric-Spheroid}.

The polar ring galaxies are unique objects for investigation of the DM halo shape  \citep{Combes-2006!Polar-Ring}.
\citet{Whitmore-etal-1987!halo-polar-ring} analysed the polar ring kinematics in several galaxies and found significant deviation of the DM distribution from the
spherical shape:  $ c / a \sim 0.5 $~\citep{Sackett-Pogge-1995!Polar-ring}, $ c / a \sim 0.3-0.4 $
\citep{Sackett-etal-1994!Polar-Ring}.
A prolate shape of the dark matter haloes was inferred by \citet{Iodice} based on the analysis of location
of the polar ring galaxies on the Tully-Fisher diagram.

Dynamics of decaying satellites also helps to constrain the dark halo properties, such as its shape, orientation, mass distribution  \citep{Ibataetal2001b, Helmi-2004!Is-dark-halo-our-Galaxy-spherical, Fellhaueretal2006}. According to \citet{Sesar-etal-2011!Halo-MW}, SDSS data suggest that
the Milky Way's dark halo is oblate ($s=0.7$) and non-axisymmetric in the disc plane ($q=0.98$) within 28 kpc of the centre.

An assumption of the axissymmetric potential for the Milky Way halo applied to the Sgr dSph tidal stream  yields controversial results: while the leading arm of the stream requires an oblate halo, the radial velocity distribution of stars in the arm agrees better with a prolate halo~\citep{VivasZinn2005}. Note that the introduction of a triaxial potential  helps to resolve this ambiguity \citep{Law-etal-2009!Triaxial-MWay-Dark-Halo-Sagittarius-Stream}.

Statistical analysis of shapes of the spiral galaxies also reveals some deviation from the disc axial symmetry, which in turn may evidence the halo triaxiality. According to ~\citet{Ryden-2006!Shape-Spiral-Galaxies}, the disc ellipticity is $0.02 - 0.3 $ in the $K_s$ photometric band. The ellipticity varies in dependence of the photometric band and of the galaxy morphological type~\citep{Ryden-2006!Shape-Spiral-Galaxies}.

Despite direct and non-direct  evidences of the halo triaxiality mentioned above,
its shape and ellipticity in specific galaxies remains poorly known. Understanding of the role that the asymmetry of
the halo potential can play in dynamical evolution of stellar discs in galaxies remains incomplete.
\citet{El-Zant-Habler-1998!Triax-halo-disk,
Ideta-Hozumi-2000!Bar-Dissolution-Prolate-Halos,
Tutukov-Fedorova-2006!asymmetry-dark-haloes,
BerentzenShlosman2006,
Hayashi-Navarro-2006!kinematics-disc-triaxial-halo,
Hayashi-etal-2007!shape-gravitational-potential-halo} studied
different aspects of how a non-axisymmetric
gravitational potential affects the galactic disc's dynamics.
The triaxial halo allows to explain the observed shape of
rotation curves in low surface brightness (LSB)
galaxies~\citep{Hayashi-Navarro-2006!kinematics-disc-triaxial-halo,Widrow-2008!Disk-Galaxies-Triaxial-Halos}.

Galaxies with warped discs were successfully
reproduced within the models with triaxial dark matter (DM) halo
\citep{Dubinski-Chakrabarty-2009!Warps-Triaxial-Halo,Roskar2010}.
The effects of the bar response to the triaxial halo perturbations and bar
dissolution in prolate halo were studied by \citet{Machado-Athanassoula-2010!bar-triax-halo} and
\citet{Kazantzidis-etal-2010!Sphericalization-Halos-Disks}.
Numerous $N$-body simulations are usually  used to study the formation and evolution of stellar bars embedded in triaxial dark matter haloes~\citep{Athanassoula2002,Berentzen2006, Dubinski2009}. The triaxial halo can destroy stellar bars due to the angular momentum exchange~\citep{BerentzenShlosman2006}. The decreasing of the bar pattern speed was demonstrated in models with triaxial dark
matter halo~\citep{CeverinoKlypin2007}.

\citet{Bekki-Freeman-2002!Spiral-Structure-Triaxial-Halo} explained the spiral structure  extended  far beyond optical limits in a blue compact dwarf galaxy NGC 2915  by the presence of a triaxial dark halo. \citet{Masset-Bureau-2003!Spiral-Structure-NGC2915-Triax-Halo} modeled the spiral structure in the outer region of the gaseous disc of NGC 2915 generated by a triaxial dark matter potential. The main result of these papers is that triaxial DM halo is able to produce a two-arm spiral pattern in external regions beyond the optical radius.  \citet{TinkerRyden2002} investigated effects of rotating triaxial haloes in the disc galaxies using $N$-body simulations with the rigid halo potential and found that the average pitch angle of the spiral structure strongly anti-correlates with the total mass of the system (disc+halo).
 \citet{Dubinski-Chakrabarty-2009!Warps-Triaxial-Halo} applied $N$-body simulations to demonstrate the formation of warps and spirals in isolated discs  embedded in triaxial haloes. They showed  that the spiral pattern may emerge in gravitational stable stellar disc through  interactions with a non-axisymmetric halo.
 Note that the evolution of the grand design spiral structure in stellar discs embedded in the triaxial dark matter haloes remains poorly studied.

In our previous paper~\citep{Khoperskov-2012!Halo-gas-spiral} we studied the dynamics of a gaseous galactic disc in a non-axisymmetric gravitational potential in general. In the present paper we perform $N$-body modelling (with $N \sim 10^6-10^7$) of stellar collisionless discs embedded in a non-axisymmetric dark halo
with focusing on the formation of the spiral pattern, its morphology and kinematics.
In Section 2 we describe our modelling technique and the initial conditions in the our galactic discs. The spiral structure formation and the kinematics of the density waves are analysed in Section 3. Discussion and main conclusions are given in Sections 4 and 5, respectively.

\begin{figure}\label{fig02}
  \includegraphics[width=1\hsize]{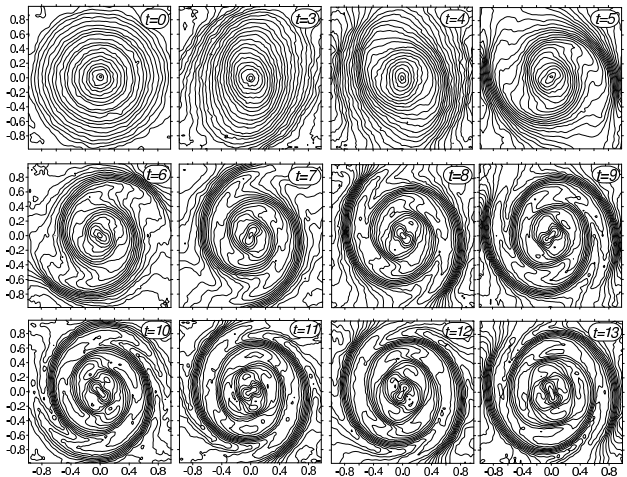}
  \caption {The surface density isolines $ \sigma (r, \varphi) $ at different time $t$ in the model with $\varepsilon = 0.1$ ($\sigma$ is shown in the log scale).
 One period of rotation at the disc's periphery corresponds to about 3 our dimensionless units.}
\end{figure}

\section{Modelling collisionless stellar discs}

Throughout this paper we study the dynamics of the galactic discs in the frames of a self-consistent $N$-body problem by solving the motion
equations for equal mass  test particles:
\begin{equation}\label{eh-ii-newton}
    \frac{d^2{\textbf{r}}_j}{dt^2} =  - m\,\sum_{\stackrel{i=1}{i\ne
    j}}^{N} \frac{\left( {\textbf{r}}_j - {\textbf{r}}_i \right)}{\left| {\textbf{r}}_j - {\textbf{r}}_i
    \right|^3} - \frac{\partial \Psi_b}{\partial{\textbf{r}}_j} - \frac{\partial
    \Psi_h}{\partial{\textbf{r}}_j} \,, \;\;\; j=1,2,...,N\,,
\end{equation}
here $\textbf{r}_j = \{x,y,z\}$ is the coordinate vector of $j$~-th particle, $M_d$ is the total stellar disc mass, $\Psi_b(x,y,z)$ is the bulge potential, $\Psi_h(x,y,z,t)$ is the halo potential and $m=M_d/N$ is the mass of a test particle.

The force of gravity in the disc is determined by the summation over all particles in Equation~(2) and calculations via the
TreeCode Top Down algorithm~\citep{BarnesHut1986}. Initial distribution functions in the equilibrium stellar disc are described in detail in our previous papers~\citep{Khoperskov-Zasov-Tiurina-2003!GravitInstab, Khoperskov-etal-2010!z-str}. The schemes of the 2nd and 4th orders we apply
provide a good accuracy of integration of Equations~(2).

We introduce the following dimensionless model units:
Our spatial distance unit is assumed to be $10$ kpc, the gravitational constant is chosen to be $G = 1$, the time unit is $91$ Myr, the velocity is expressed
in the units of $110$ km s$^{-1}$. Assumptions $G=1$, $V=1$ lead to the total mass of model galaxy equal to about $10^{11} M_\odot$ within $R=1$.

The number of particles in our models is $ N = (1-10) \cdot 10 ^ 6 $.  The potential for $i$-th particle at $j$-th point is $\Psi_{ij}=-m/\sqrt{r_{ij}^2+r_c^2}$, where the cut off radius is $ r_c \leq 10^{-3}$ (which corresponds to $\leq $ 10 \, pc). The opening angle parameter $\Theta$ is assumed to be  $0.1-0.5$, although we did not find significant variations of the results between models with different $\Theta$. The results presented in this paper were obtained assuming $\Theta = 0.1$,
if alternative value is not specified. This values of $r_c$ and $\Theta$ determine the accuracy of the calculations. The integration step is chosen to be $2\cdot 10^{-3}$, which approximately corresponds to $2\cdot 10^5$ yr.

\begin{figure*}
  \includegraphics[width=0.33\hsize]{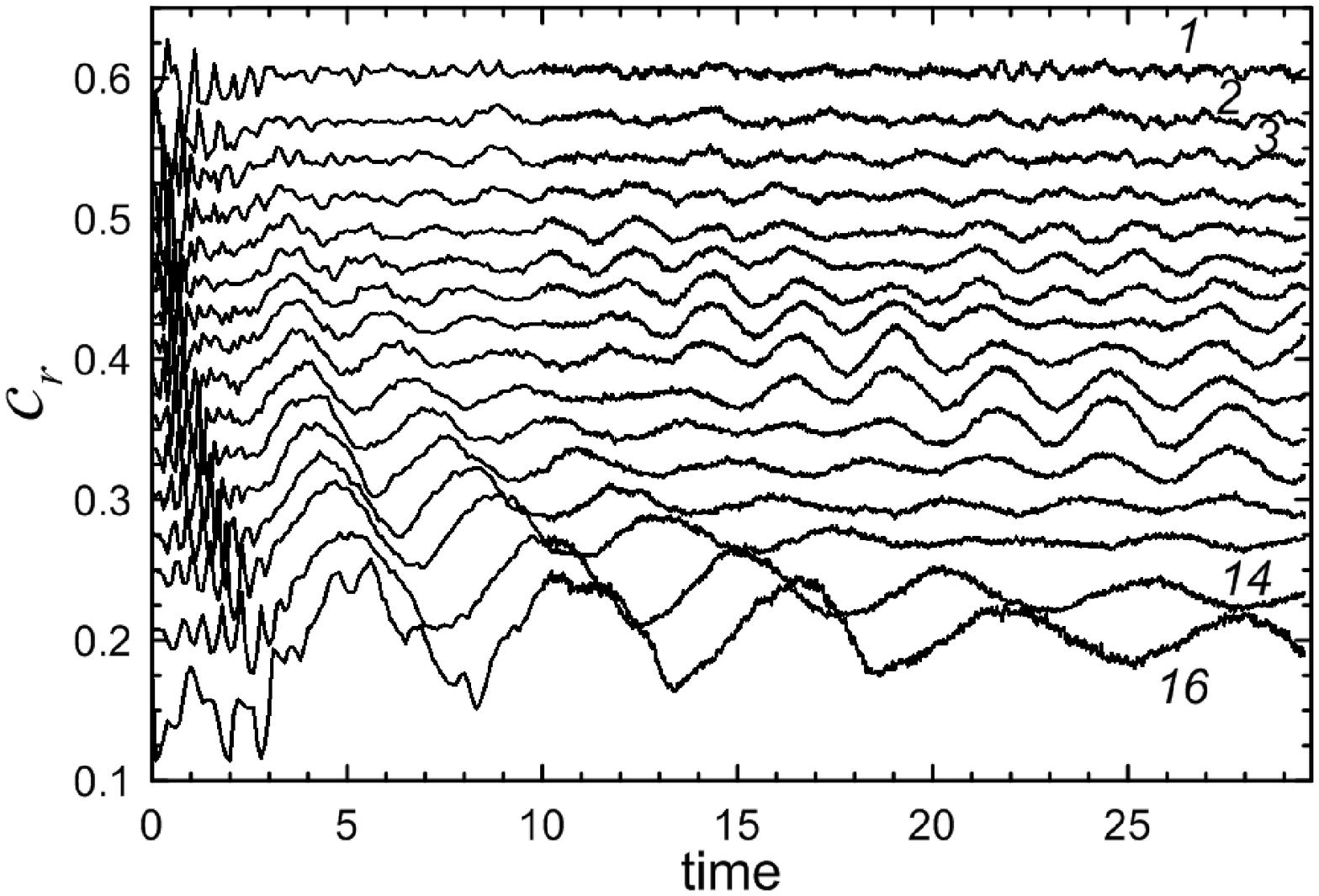}
  \includegraphics[width=0.33\hsize]{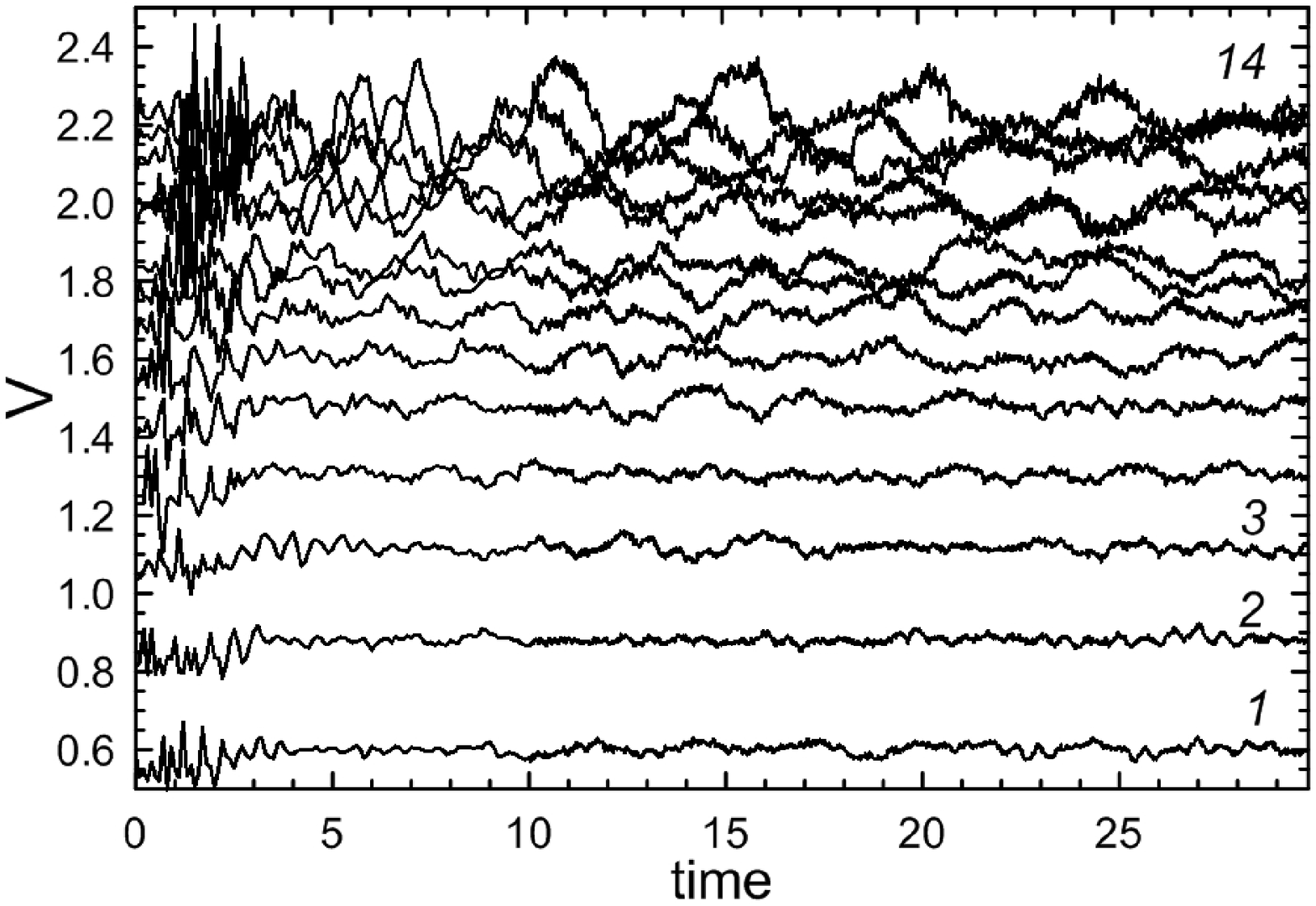}
  \includegraphics[width=0.33\hsize]{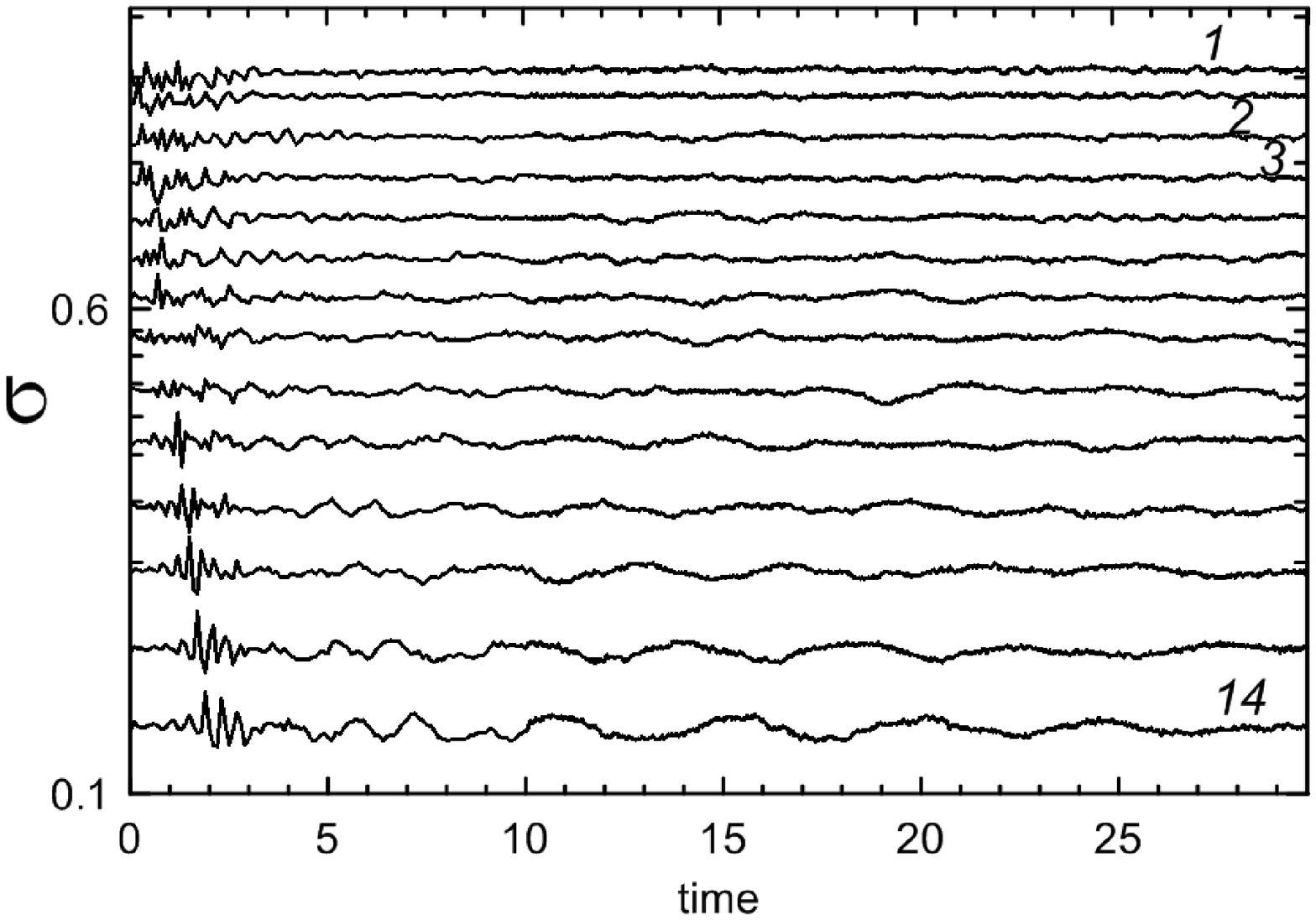}
  \caption{Evolution of the disc  parameters averaged over the azimuthal angle at various distances from the centre~(curve number increases from the centre towards the periphery). Strong fluctuations at the beginning of the experiment ($t<5$) are due to the deviation from precise balance at the initial moment.}\label{fig03}
\end{figure*}

Basic model parameters are constrained in the range suggested by observations as follows:
   \begin{itemize}
    \item Relative dark halo mass $1\le\mu=M_h/M_d\le 4$ within the limits of the stellar disc  $R=4r_d=1$, where $r_d$ is the radial scalelength of the disc.

      \item $\mu_b=M_b/M_d=0 - 0.3$  is the relative stellar mass of the bulge with scalelength $r_b=0.1 - 0.3$.

        \item $\tau_0 \approx 3$ is the period of rotation at the periphery of the disc.

    \item Nonaxissymmetric halo parameter $\varepsilon=1-b/a=0 - 0.25$ in the disc  plane~\footnote{It should be emphasised that the value $\varepsilon = 1 - b/a$ calculated for the dark halo volume density distribution differs from that for the gravitational potential $ \varepsilon_\Psi $. Approximately, $ \varepsilon_\Psi \simeq 3 \varepsilon$ \citep{Binney-Tremaine-2008}.}.

    \item $\Omega_h=0 - 0.3$ is the angular rotation velocity of the halo. The halo corotation radius is located far outside the stellar disc~($\Omega_h=0.3$ or about $3$ km s$^{-1}$ kpc$^{-1}$).

     \item $\tau_h=0 - 20$ is the time scale of variation of the non-axisymmetry parameter from 0 to $\varepsilon$.

     \end{itemize}

We assume the exponential surface density distribution in the disc as an initial condition:
   $ \sigma (r) = \sigma_0 \exp(-r/r_d) $, where $ r_d = 0.25 $  of dimensionless distance units is the radial scale length in the disc. At the initial stage, all stellar disc particles are situated within the radius $ r \le 5r_d = 1.25 $. We assume a flat rotation curve with the maximum  velocity $V_{\max}\simeq 2$ in dimensionless units, which corresponds to the rotation period of $\tau_d\simeq 3$ at the disc's periphery.

Following \citet{Begeman-etal-1991!rotation-curves-Dark-haloes}, we use a pseudo-isothermal model of the density distribution in the halo:
 \begin{equation}\label{eq-rho-halo-iso}
    \varrho_h(r) = \frac{\varrho_{h0}}{1 + (r/a)^2} \,.
\end{equation}
 This distribution of the dark mass provides a constant
circular velocity $V_c$ for $r\gg a$. The gravitational potential can be written
 \begin{equation}\label{Eq-Elista-potent-halo}
    \Psi_h(x,y,z) = 4\pi G \varrho_{h0} a^2\cdot \left\{\ln(\xi) + \frac{\rm arctg(\xi)}{\xi} + \frac{1}{2} \ln \frac{1+\xi^2}{\xi^2} \right\}
\end{equation}
\citep{Khoperskov-2012!Halo-gas-spiral}, which corresponds to the density law (\ref{eq-rho-halo-iso}) for the case of $ a = b = c $.

Note that the cosmological simulations predict that virialised dark matter haloes should have a universal density profile. For a spherically averaged density distribution it can be written  \citep{Navarro-etal-1997!Model-halo}:
\begin{equation}\label{eq-rho-halo-NFW}
    \varrho_h(r) = \frac{\varrho_{h1}}{\left( r/a \right)\left( 1 + (r/a)\right)^2} \,.
\end{equation}
Its distinctive feature is the inner density cusp, which however may be dissolved due to its interaction with baryonic matter in the central region of galaxies  (see f.e. \citet{Gustafsson2006,Maccio2012,KhoperskovShustov2012}). Moreover, inevitable singularity  emerges at $r=0$ in the case of transformation from a spherical NFW profile to the triaxial shape. Another widely used dark matter density distribution is Burkert's profile \citep{Burkert-1995!Structure-Dark-Matter-Halos}, which usually fits observing data very well~\citep{Gentile2005}. It is very close to the NFW profile: it has a cored density distribution in the inner part~($r<a$). Several more halo density profiles were introduced, such as exponential profile~\citet{Fux-1997!MW-stellar} and Einasto profile~\citep{Einasto1965}.
All density distribution functions mentioned above (except~(\ref{eq-rho-halo-NFW})) allow generalization of the spherical dark matter distribution shape to the triaxial case using the formal replacement $ r \rightarrow \xi $ (see eq.~\ref{eq-shape-halo}). Note however, that the choice of the accepted law for $ \varrho_h (r) $ makes only a small effect on the results described below.

Rotation pattern in the dark haloes predicted by $\Lambda CDM$ cosmological models is similar to that in giant elliptical galaxies, where the kinetic energy of the rotation is much smaller than that in the stellar chaotic motions.  The origin and properties of angular momentum of DM haloes were investigated by \citet{Vitvitska2002,Maccio2007,Bett2007}. For the disc  galaxies cosmological simulations suggest slow rotation of their dark matter haloes within $ R_{opt} $.  The  rotation (tumbling) of the overall shape of the halo in
$\Lambda$CDM  simulations was discussed by~\citet{BailinSteinmetz2004,BryanCress2007}. Authors conclude that the tumbling rates of the
haloes are less than one km s$^{-1}$ kpc$^{-1}$. Evidently, a disc subsystem rotates much faster than a halo. Hereafter we use a non-rotating or slow rigidly rotating model halo with the angular velocity $\Omega_h $, which places the
corotation radius at the disc's periphery $ \Omega_h \simlt V_ {\max} / R_{opt} $.
 Our principal results are shown for the case of rigid DM halo, while the situation with  the live halo is considered in Section 3.4.

\section{Disc dynamics: properties of the spiral structure}

Propagation of the density waves of gravitational nature in a self-gravitating stellar disc requires that the stellar velocity dispersion were close to the stability condition $Q_T \leq 2$ (or $Q_T \leq 3$ at the periphery of the disc, see~\citet{Khoperskov2003}), where $Q_T$ is the Toomre stability parameter~\citep{Toomre-1964!Criterion-Toomre}. The case $Q_T = 1$ is unstable due to non-radial perturbations. In the following section we consider gravitationally stable discs ($Q_T \sim 2-4$ at $r \sim (1-2) \cdot r_d$) and study the pure effects from the non-axisymmetric dark matter distribution. We also compare the properties of the spiral pattern
generated by triaxial haloes in the cases of stable and unstable discs (see Section 3.3).

\subsection{ Dynamically stable stellar discs}

The initial distributions of the radial $c_r$, azimuthal $c_\phi$ and vertical velocities dispersions $c_z$ are chosen to be high enough to prevent the disc  gravitational instability, which is able to generate spiral waves in the initially axisymmetric model by its own.
The stellar velocity dispersion should be high enough to support the gravitational stability against perturbations in the disc plane and against the bending instability
\citep{Khoperskov-etal-2010!z-str}. An iterative procedure is applied for making marginally stable 3-D stellar discs as described by~\citet{Khoperskov-Zasov-Tiurina-2003!GravitInstab}.

First, we tested  the disc's  stability in the potential of the spherical halo. We assureed that  the initial disc  parameters ($c_r(r)$, $c_\varphi(r)$, $c_z(r)$, rotation velocity $ V(r) $, surface density $\sigma(r)$, and the disc  vertical scale $ h(r) $) do not change significantly during several revolutions of the disc. As an example, figure~1 illustrates  the temporal variation of the $c_r(r)$. Collisionless stellar disc  in this model does not undergo any dynamic heating neither  to binary particle interactions nor due to collective processes. Under the chosen conditions the disc  remains axisymmetric during dozens revolutions (which corresponds to a few billion years).

\subsection{Generation of the spiral structure in the stellar disc}

Our simulations started from the spherical halo and axisymmetric stellar disc, then we slowly and adiabaticically (over time scale $ \tau_h \sim 3$) increased the halo's non-axisymmetric parameter $\varepsilon$.

\begin{figure}\label{fig04}
\includegraphics[width=0.9\hsize]{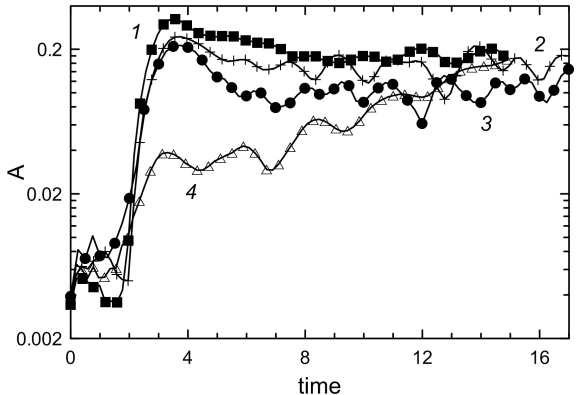}
  \caption{Amplitudes of the Fourier harmonics for $m=2$ mode for the entire disc  in different models: \textit{1} : $\mu=4$, \textit{2} : $\mu=3$,
 \textit{3}: $\mu=2$,
 \textit{4}: $\mu=1$~($\varepsilon=0.15$, $r_b=0.3$ for all models). }
\end{figure}

\begin{figure}\label{fig05}
\includegraphics[width=0.9\hsize]{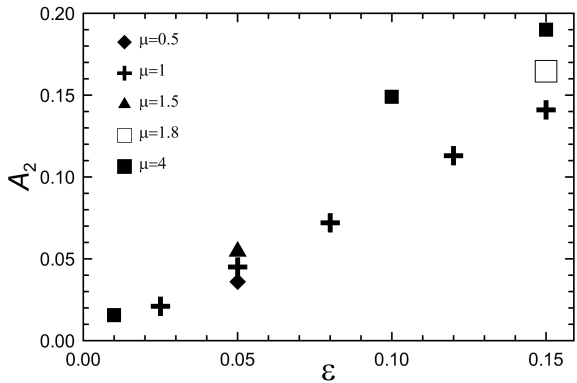}
  \caption{Mean amplitude of the time-averaged Fourier harmonics for the entire disc in the two-arm mode as a function of the parameter $\varepsilon$.}
\end{figure}

We performed about 30 modelling runs with different sets of initial model parameters mentioned above. It is  evident that in all models with $\varepsilon >0.05$ the  presence of the non-axisymmetric dark halo excites the formation of a two-arm spiral pattern in the stellar disc. In figure~2 we show snapshots of the typical evolution of the surface density in the model with $ \varepsilon = 0.2 $. Within 1-2 periods of rotation $\tau_0$, a two-armed spiral structure forms in the initially axisymmetric exponential disc. A nested $\Theta$-like structures and a central bar can be identified in the inner disc  region ($ r \leq 0.2 \leq r_d=0.25 $).
The rigid triaxial halo models (models without the disc feedback to the halo) generate a long lasting global spiral pattern that extends from the centre to the disc periphery. It is important to note that interactions between the disc and non-axisymmetric dark halo's potential do not introduce any significant systematic variations to the azimuthally-averaged disc  parameters.

Evolution of $c_r(r)$, $V(r)$ and $\sigma(r)$ is shown in figure~3. Note that the centre of the disc (curves 1 and 2) remains unperturbed over the course of the simulations, whereas quasi-periodic radial oscillations can be observed, especially in the outer part of the disc. In contrast to the global gravitational stochastic  spiral structure that develops in the slightly unstable discs~\citep{DeSimone2004}, here the heating of  the stellar disc is insignificant ~(except for the short initial stage of the experiment).

The model parameters $ \mu $, $ \mu_b $, $ r_b $ and $ \tau_h $ significantly affect the growth increment of the spiral structure. The amplitude of the stellar density wave at early stages in the simulations is primarily determined by the relative halo mass parameter $\mu$ (see figure~4). However, after several $\tau_0$, two-arm spirals reach the similar amplitudes at a fixed value of $\varepsilon$, independently of $\mu$.
 Our modelling demonstrates the dependence of the eventual amplitude $A_2$ of the spiral waves on the triaxiality parameter $\varepsilon$ for the spiral mode $m=2$~(see figure 5). It seems that  $\varepsilon$ is a crucial parameter, which defines the properties of the strength of the spiral structure as a whole. Despite the fact that the growth rate of the spiral waves depends on the relative mass of the halo~(see figure~4), the final amplitude of the waves is determined by the non-axisymmetric parameter $\varepsilon$~(see figure~5).

There are two distinctive properties of the spiral structure generated by the triaxial halo. First, the spirals are generated in a sufficiently hot (i.e. gravitationally stable) disc with the Toomre parameter $Q_T \sim 2-4$, in contrast to the standard density waves theory. Therefore, this mechanism can produce spirals even in the over-stable stellar discs, e.g. in the outer regions of galaxies far beyond their optical size limits, where the disc surface density is much lower than the gravitational stability threshold.  Second, the formation of the spiral structure leads to a weak stellar disc warming-up during the initial stage of the spiral pattern development, and after that the velocity dispersion components only slightly increases even in the case of high amplitude spiral waves~(figure~3). These properties are significantly different from those seen in $N$-body simulations of gravitationally unstable discs, in which the development of the gravitational instability is accompanied by significant growth of  the velocity dispersion during one period of rotation. It should be noted that a slight disc heating for large amplitude waves that we observe in the our modelling  may be a result of limited accuracy in the numerical solution of the motion equations (2).
Indeed, we found that the disc heating decreases with the increasing of the number of test particles.
The heating up effect may also be attributed to the lack of perfect disc equilibrium at the beginning of  the simulations.
The final conclusion about the low  disc heating requires modelling with higher spatial resolution.

\subsection{Simulations with different stellar disc initial state}
While we started simulations with the initially stable disc in the previous section, we should consider
whether the initial gravitational instability of the disc can qualitatively change the results of section 3.2. To reveal the effect of self-gravity  we have performed simulations for initially unstable disc by assuming the relative halo mass $M_t /M_d =1$ and $Q_t=1$. We started with the axisymmetric halo to be sure that the evolution of the disc and spiral structure is fully determined by the global gravitational instability of the entire disc, and that the formation of the spiral waves is accompanied by the disc heating up to the marginal stability condition (see Figure 6).  After that we ran another  simulation with the same initial conditions, but
assuming the halo with $\epsilon = 0.1$.  Relatively low halo mass led to slower growth of the wave amplitude than that shown in Figure 4 for $\mu =1$.
Curiously, the final amplitude of the spiral waves is practically the same independently on the mechanism of their generation (either non-axisymmetric halo +gravitational instability, or pure gravitational instability). A general character of the evolution of the velocity dispersion is also very similar in both models. It confirms that the non-axisymmetric halo does not contribute essentually to the eventual disc heating.

To  analyze the direct  role of a triaxial halo and to separate it from effects of the disc initial state, below we compare different numerical models of evolution of  gravitationally stable stellar discs in the halo field. It was assumed that the disc always remains in a quasi-equilibrium state even it is non-axisymmetric, since the halo shape changes slowly and adiabatically in our models. Model (\textit{a}) is a ''reference'' model for the initially round-shaped disc  described in Section 3.2. In model (\textit{b}) we introduce a very large timescale for the halo triaxiality growth ($\tau_h=15$) for initially round disc to be sure that the disc  has enough time to follow the slowly changing halo potential. In our third  model (\textit{c}) we use the initially elliptical disc with the same spatial major axis orientation as for the non-axisymmetric halo.  In our last model (\textit{d}) we took the results  of long-time evolution of a ''reference'' model of round disc in the non-axisymmetric halo at $t=20$. After that the spiral structure was radially  smoothed  (which introduced slight ellipticity to the disc), and next simulation was started from this equilibrated, spiral-free initial state.

\begin{figure}\label{fig057}
\includegraphics[width=0.9\hsize]{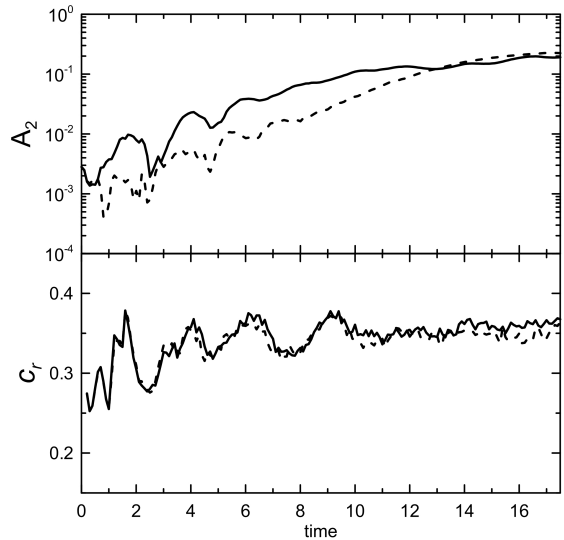}
  \caption{Top panel: evolution of the Fourier harmonic $m=2$ in a gravitationally unstable disc ($Q_T = 1$) embedded in a non-axisymmetric halo $\varepsilon=0.1$ (solid line) and in a symmetric halo (dashed line). Bottom panel: evolution of the velocity dispersion in the same two models at the radius of $r=0.5$.}
\end{figure}

Results of modelling are shown in Figure 7. In all four models the spiral structure emerges. The main difference between the models is observed at the central part of the disc. In models (\textit{a}) and (\textit{b}) a $\Theta$-like structure forms. In the model (\textit{c}) of initially elliptical disc a large bar was formed, and beyond the bar there is a tightly wound spiral structure. In the case (\textit{d}) there is a small bar and very even spiral arms. Formation of a bar in the potential of triaxial DM halo links our models with those developed by \citet{Berentzen2006}, where the bar generation and bar-halo  interaction are also described. It is essential that our models demonstrate the emergence of the spiral arms in the very outer regions of the disc driven by the halo triaxiality, independently of the disc gravitational stability and its initial shape.

\begin{figure}\label{fig056}
\includegraphics[width=1\hsize]{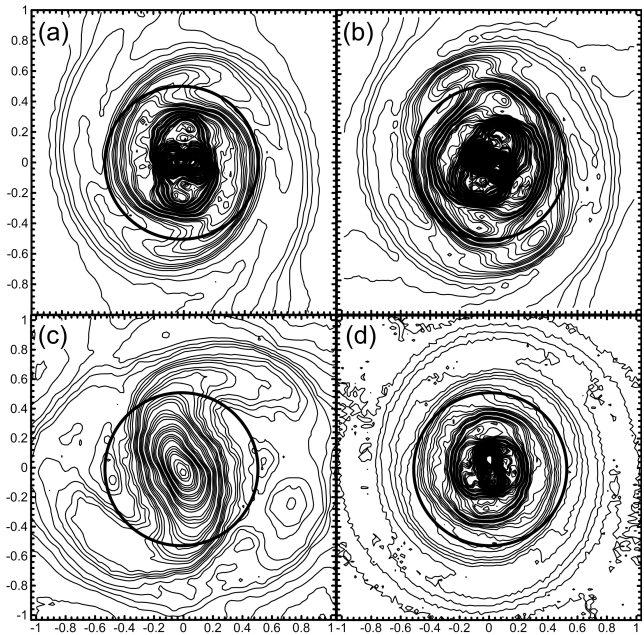}
  \caption{Disc surface density at $t=8$ in models with different initial state of the stellar disc for the non-axisymmetrical halo with $\varepsilon=0.1$: (a) the reference model of the symmetric disc; (b) evolution with long term quasi adiabatic growth of $\varepsilon$ with $\tau_h=20$; (c) the model with a non-axisymmetric disc; (d) the model started with a radially smoothed disc made with those from the model (\textit{a}) after $t=20$ evolution.}
\end{figure}

To analyze the properties of the spiral structure without considering the central disc regions, we considered the evolution of the Fourier harmonics and velocity dispersion beyond the radius $r = 0.5$ (the solid circle in Figure 7). Despite the difference in the disc initial state, the amplitude and its time-dependent evolution are rather similar (see Figure 8, top panel). The final amplitude of the spiral waves is about 10\% in all cases. The initial values of the Fourier harmonics
$m=2$ in (\textit{c}) and (\textit{d}) models are large because of the initial ellipticity. The model (\textit{b}) with a long timescale of the shape formation  of adiabatic halo shape transformation  reveals a slow quasi-steady increasing of the spiral wave amplitude, which follows the slow growth of the halo triaxiality.
These numerical experiments confirm that formation of the spiral structure in the models is entirely due to interaction of the disc with the non-axisymmetric halo. After the spirals have formed, there is no significant dynamical heating introduced to the disc (see bottom panel in Figure 8). Some initial growth of the velocity dispersion observed in the model (c) is caused by the relaxation processes in the elliptical-shaped disc while it transits  to another (also elliptical) disc state due to interaction with equally oriented halo. After this transition has finished the disc heating is ceases.

\begin{figure}\label{fig058}
\includegraphics[width=0.9\hsize]{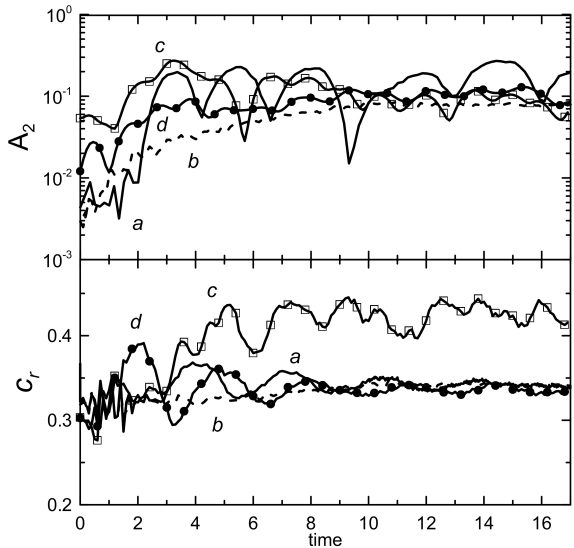}
  \caption{Analysis of the spiral arms in different models beyond $r=0.5$ --- the designations are kept the same as in figure \ref{fig056}.
  Top panel: evolution of the Fourier harmonic $m=2$.
  Bottom panel: evolution of the velocity dispersion.}
\end{figure}

\subsection{Kinematics of the spiral structure}
The non-stationary evolution of the spiral waves is characterised
by quasi-periodic oscillations that are caused by the radial propagation of spiral waves~(figure~3). The spiral wave pattern rotates with the angular velocity~$ \Omega_p$, which slightly changes with time.  Figure~9 shows the evolution of the wave amplitude at different radii. Estimation of the phase shift along the radius enables us to find the pitch angle of the spiral pattern, which ranges from $5^\circ$ to  $30^\circ$. For illustration,  we assume it to be  $\sim 12^\circ$ in our model with $\varepsilon = 0.1$. In our simulations, the pitch angle is a function of radius: it grow from the centre to the periphery in the galactic disc, so the spiral arms become more open. This radial dependence may be used for the revealing of the halo non-axisymetry signature in the real galaxies.

A distinctive feature that we observe in our modelling is that the spiral pattern does not show a rigid-body rotation.
Instead, at any radial distance in our models, its possesses a two-pattern angular velocity, which switches from $ \Omega_{p1} $ to~$ \Omega_{p2} $ and back~(see figure~10) twice during the orbital period.  Most of the time the wave rotates more slowly (angular velocity ~$\Omega_{p1}$), which corresponds to the
case when the corotation radius $ r_{c1} $ locates at the periphery of the disc or beyond it.
When the wave passes through the dark halo potential well, its angular rotation rises to $ \Omega_{p2}> \Omega_{p1}$,
and the corotation radius $ r _{c2} $ moves inward down to~$(2-3) \cdot r_d $.

\begin{figure}\label{fig06}
\includegraphics[height=0.57\hsize]{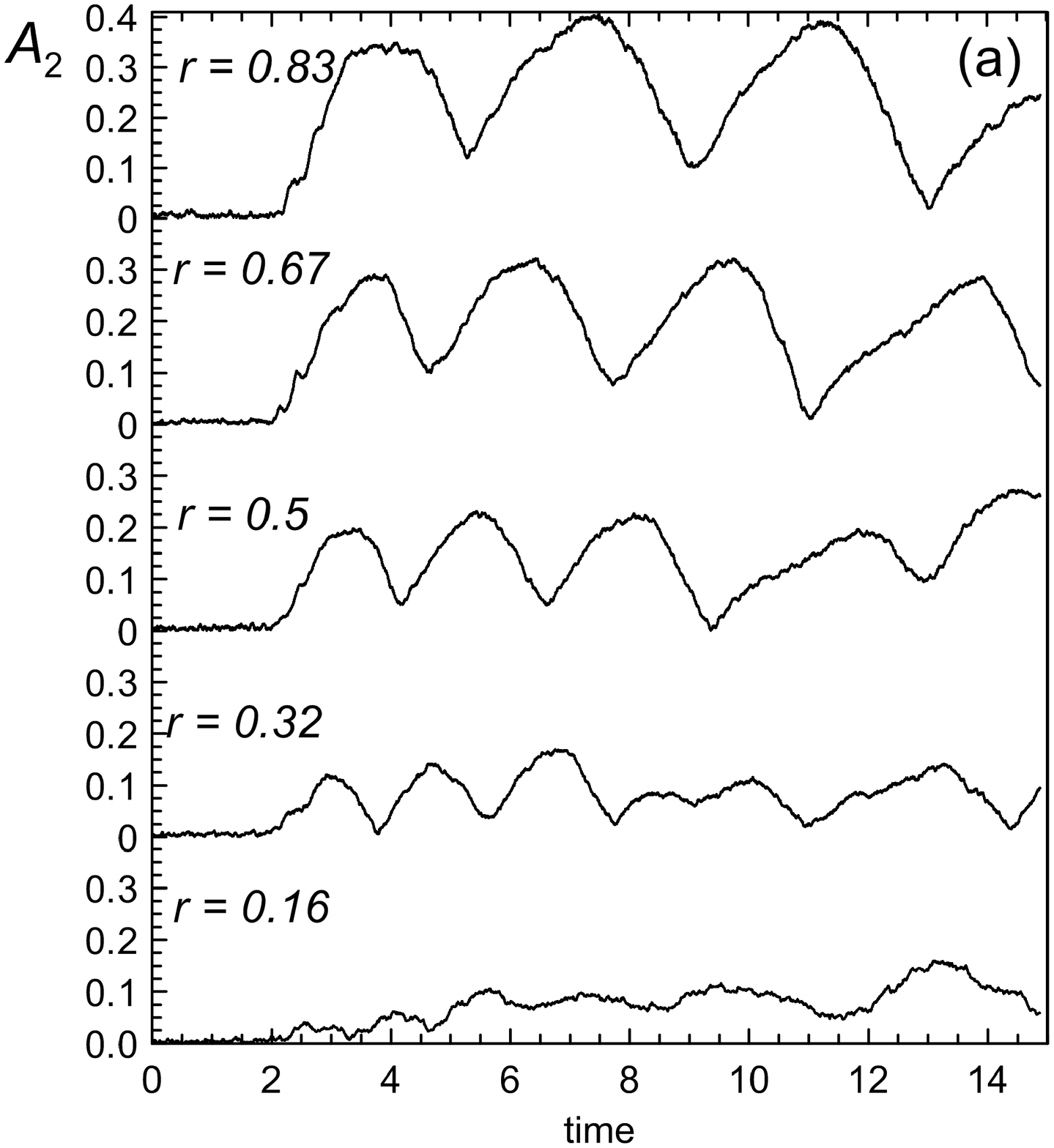}\includegraphics[height=0.57\hsize]{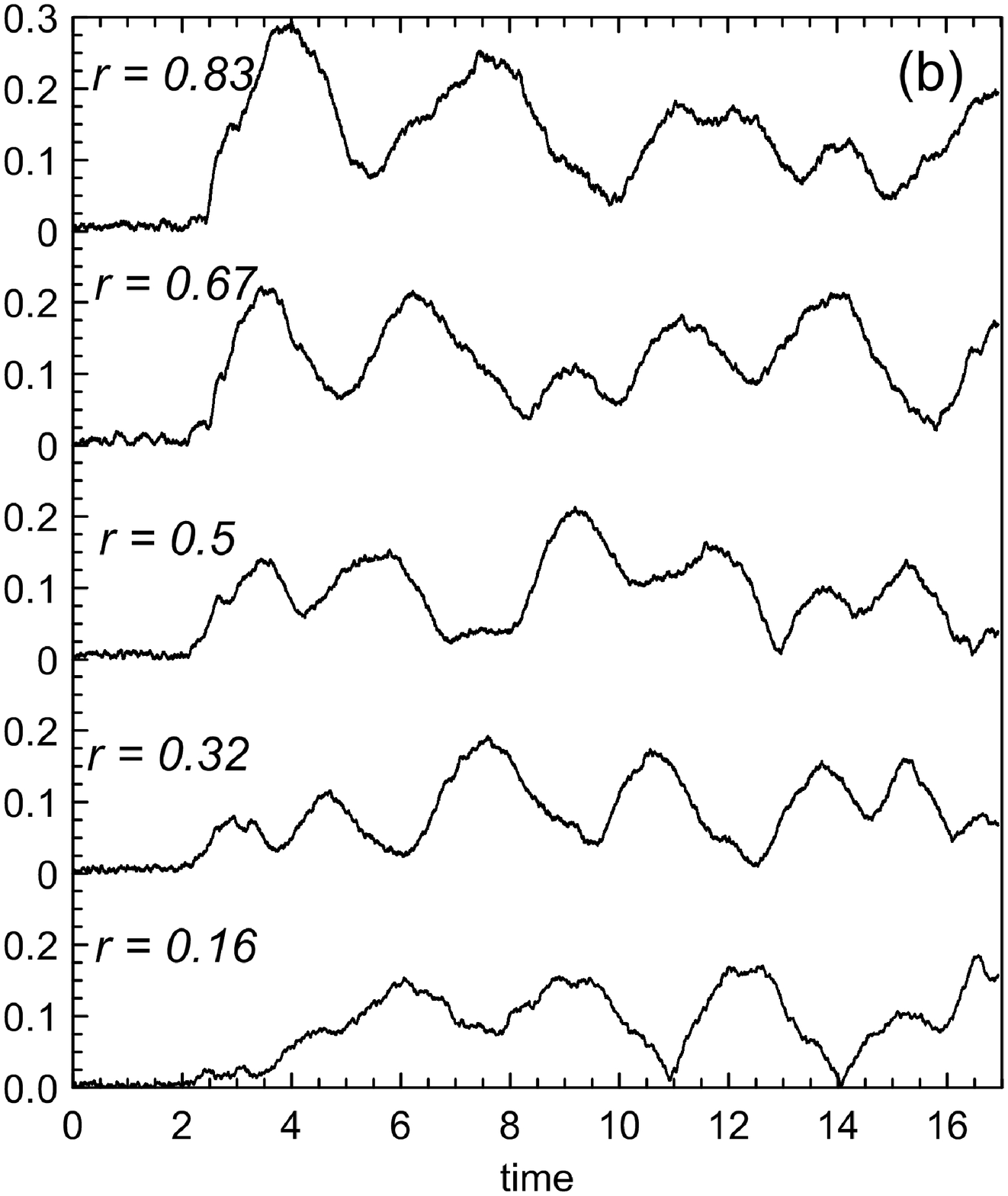}
\caption {Evolution of the amplitude of the Fourier harmonic $ m = 2$ in the surface density distribution
at various distances from the centre of the disc $ r $ as a function of
time: \textit{a}) $ \mu = 4 $ and $ a = 0.3 $; \textit{b}) $ \mu = 1.8 $ and $ a = 0.3 $. }
\end{figure}

\begin{figure}\label{fig07}
\includegraphics[width=0.9\hsize]{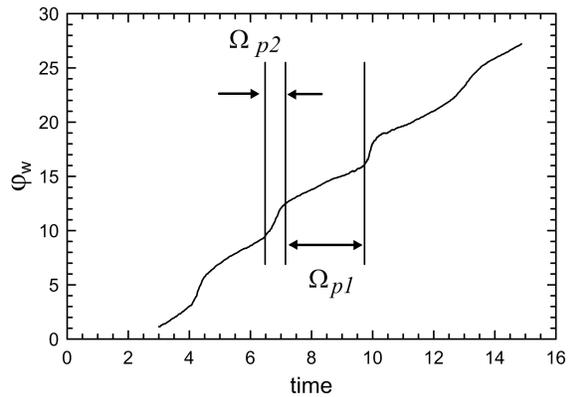}
\caption {The change of the wave phase $ \varphi_\textrm{w}(t) $ at fixed
  radius $ r = 0.67 $ (see figure~9a). There are two stages distinguished by their different angular velocities
   $ \Omega_p = d \varphi_\textrm{w} / dt $: slow rotation with $ \Omega_ {p1} $ and a short phase of fast rotation with $ \Omega_{p2}> \Omega_{p1} $}
\end{figure}

\begin{figure}\label{fig0511}
\includegraphics[width=0.9\hsize]{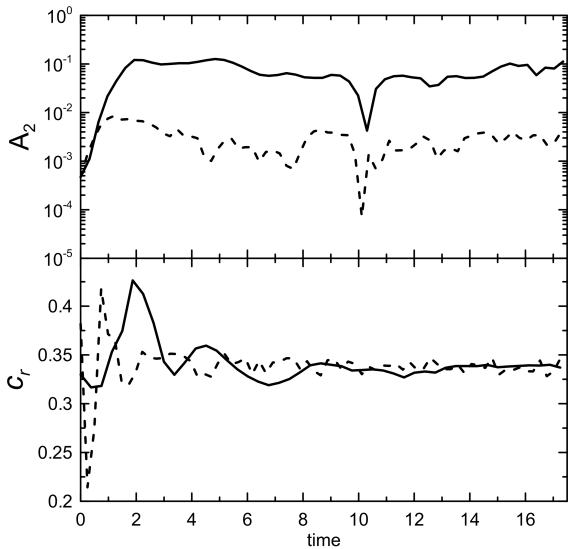}
  \caption{Comparison of the disc evolution in the models with live halo: the solid line is the case of a non-axisymmetric halo $\varepsilon=0.1$, the dashed line is
  a case of symmetric halo $\varepsilon=0$. Top panel: evolution of the Fourier harmonics $m=2$. Bottom panel: evolution of the velocity dispersion.}
\end{figure}

\subsection{Models with the live halo}
To analyse the influence of possible interior substructure in the dark matter, e.g. of presence of subhaloes predicted by the standard cosmology theory~\citep{Klypin1999, Bullock2001,Rocha2012}, we ran our  $N$-body simulations of interaction between the triaxial live halo and the stellar collisionless disc.
We performed the simulations in several steps.
First, we construct a quasi-steady triaxial live halo with a fixed stationary disc  (a "frozen disc") inside. The halo is characterised by the parameter $ \varepsilon \simeq 0.2 -  0.3$ as described before, which is initially assumed to be constant along the radius.
All disc particles were positioned  within the radius $R_{opt}=1.5$, and the halo particles were extended to  $R_h=5$. The total number of  particles
was  $4\cdot 10^6$ for the halo and $2 \cdot 10^6$ for the disc. The total halo mass in our modelling was $M_h=4$.
After several dynamical times the halo particles reached the state of relaxation and the initial numerical noise dissolved. Then we "unfreeze" the stellar disc particles instantly. Right after that, a two-arm spiral structure started forming, similar in its kinematics and morphological properties to that appeared in the case of rigid halo~(see figure~12).

\begin{figure}\label{fig08}
\includegraphics[width=1\hsize]{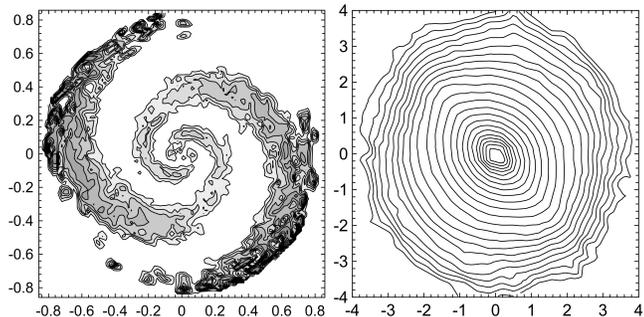}
\caption { The spiral pattern (left) and the halo isodenses in projection to the disc's plane (right) in our model with the live halo.}
\end{figure}

It is important to consider the effects of the stellar disc feedback on the dark halo shape in projection to the disc plane  for a wide range of the non-axisymmetry  parameter $\varepsilon$. In figure~12 we demonstrate the isodensity contours of the dark halo projected into  the disc's plane at $t=12$~(about 4 rotation periods at $R_{opt}$). We trace the evolution $ \varepsilon $ of the halo at three regions separately: 1) the central region $ r_1 \le r_d = 0.25 $, 2) the disc periphery at  $ r_3 \ge 1 $, and the intermediate region $ r_1 <r <r_3 $~(see figure~13). Up to time $ t = $ 10, the amplitude of the spiral perturbations remains small and there is no noticeable  stellar disc  feedback on the halo shape. Some energy initially transfers from the triaxial halo particles to the disc, causing the spiral pattern generation. After that  the central halo region becomes more axisymmetric, which means some energy is returned back to the halo due to the interaction with the spiral pattern. Local non-axisymmetric parameter in the intermediate zone also decreased down to  $ \simeq 0.15 $. The influence of spiral waves on the halo shape in the outer region remains insignificant.

We executed  several  simulations with live and symmetrical halo to investigate whether the disc and halo particle interactions can produce the disc heating. Such  dynamical interaction could also generate some transient structures. Our modelling demonstrates the lack of the dynamical disc heating, at least in the marginally stable stellar disc in the presence of the live DM halo. In early moments of the modelling some oscillations  of the disc velocity dispersion caused by relaxation of a small initial imbalance in particles distribution is observed~(see the bottom panel in figure~11). After the relaxation is finished, the increasing of the velocity dispersion stops. Transient structures forming in the disc have an insignificant amplitude $<1\%$ in the case of the symmetric halo~(see e.g. top panel in figure~11). Note that its amplitude is several times larger than in the model with rigid halo model.

\begin{figure}\label{fig09}
\includegraphics[width=0.9\hsize]{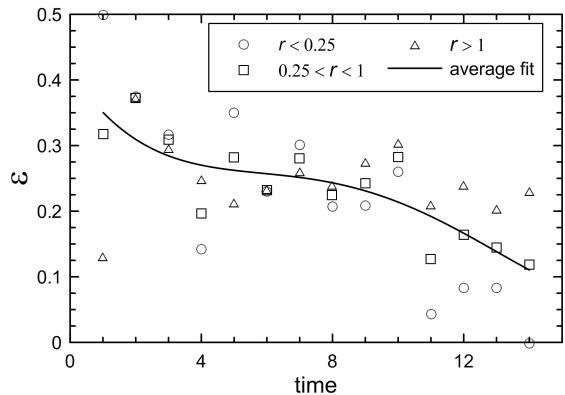}
  \caption {Evolution of the $\varepsilon$ in the live halo at different radii. The line shows the best fitting for the central part of the disc. }
\end{figure}

\section{Discussion}
Cosmological simulations show that dark matter plays significant role in formation and evolution of galaxies of all morphological types. In this paper we consider a possibility of generation of a spiral pattern in the presence of non-axisymmetric dark halo.

The key question related to the spiral structure in the observed galaxies is whether the spirals are transient \citep{Sellwood2011} or long-lived features \citep{BertinLin1996}. In this paper we bring the existing  evidences of non-spherical and, in general case, a triaxial dark matter distribution within and around the galaxies. Our numerical simulations suggest that the long-lived spiral patterns in stellar discs may be generated not only by the internal mechanisms (as it is traditionally assumed) but also via gravitational interaction between A triaxial dark halo and the galactic disc. Large-scale spirals emerge in our modelling even if the halo's triaxiality is as small as $\varepsilon \sim 0.02 - 0.05$. We demonstrate that mechanism of the long-lived spiral pattern formation discussed above acts very effectively even in dynamically hot stellar discs~(initial value of the Toomre parameter $Q_T \sim 2-4$). Our preliminary simulations shows that the presence of a gaseous component does not change significantly this picture~\citep{KhoperskovShustov2012}.  Note that similar mechanism can explain the existence  of grand design spiral pattern  observed in some dwarf early type galaxies~\citep{Jerjen2000, Lisker}: we can assume that their spherical components are slightly triaxial.

We also infer that the halo axes orientation can be found from kinematic studies of the spiral pattern. As we showed above, the pattern angular velocity cannot be described by a single constant value, as in the classical axisymmetric models: at any  radial distance  it changes  periodically twice for the  period of rotation, as a result of nonaxisymmetry of the gravitational field of the halo (see figure 10). In principle, detailed measurements of the stellar velocity field can help detect the jumps of the phase velocity in the pattern (see figure~10), which spatially should point us at the major axis of the halo. The amplitude of the angular velocity perturbations can trace the deviation from the symmetry of the halo. The best candidates to this halo diagnostics are isolated non-barred spiral galaxies with small pitch angle of their spiral arms ($\leq 20^\circ$). Apparently the present day accuracy of the methods of the $\Omega_p$ estimation makes tracing its variation along the radial and azimuthal direction a difficult problem.
Note however that there are some observational evidences of multiplicity of the pattern angular velocity in the inner and in the outer regions of galactic discs (see f.e. \citet{ButaZhang2009}). A combined estimation of the bar pattern speed in the Milky Way obtained  from the gas dynamics by \citet{Gerhard2011}  gives  $\Omega_b\simeq 52 \pm 10$~km\,s$^{-1}$\,kpc$^{-1}$  and places the corotation radius near  the end of the bar at $ 3.5-5$~kpc). On the other  hand, the Milky Way spiral arms rotate much slower: observations constrain the speed of the spiral pattern in the range of $\Omega_s\simeq 17 - 28$\,km\,s$^{-1}$\,kpc$^{-1}$ \citep{Chakrabarty2007}, what corresponds to the corotation radius position near the solar neighborhood ($\sim 8$~kpc). Thus, it is possible that the  outer spiral structure in the Milky Way is relevant mostly to the triaxial halo-disc interaction, whereas the inner spiral arms are connected with the Milky Way bar.

 We demonstrate  in figure~13 how the dark matter halo becomes more spherical in the centre due to the spiral structure formation in the case of the initially  triaxial halo. However, we bring detailed consideration of effects of the angular momentum feedback from the disc to the halo beyond the scope of this research. The dynamical interaction between the stellar disc and the DM halo occurs at resonances, and the result is the increasing of the halo rotation~\citet{Athanassoula2003}. It is worth mentioning that the angular velocity of the bar in barred galaxies decreases in time due to dynamical frictions against the dark matter halo~(see f.e.~\citet{DebattistaSellwood2000}). Numerical cosmological simulations reveal the possibility of variations of  the axes ratios in the galactic haloes during their evolution. Here we note that parallel to the secular processes  (e.g. see~\citet{Debattista2006}) the exchange by the angular momentum and energy  between the live  halo and density waves it creates  may also lead to the quasi-stationary periodical variation of  parameters of the spiral arms (including their pitch angle) with time. The latter, in turn, may cause a variation of the morphological type of a spiral galaxy during its evolution~(see \citet{MazzeiCurir2003}).

\section{Conclusions}
We conducted high resolution $N$-body simulations to study interactions between the stellar disc and triaxial massive dark halo in terms of the spiral structure formation in disc for different initial disc and halo parameters.
Our results are most relevant to the galaxies with mass-geometric properties similar to the Milky Way. The main conclusions are as follows:
\begin{itemize}
\item[1.] A triaxial dark matter halo can produce and maintain a long-lasting (at least a few Gyr) grand design spiral structure in a  stellar disc even if the disc is gravitationally stable.

\item[2.] As in the case of gaseous discs  (see~\cite{Khoperskov-2012!Halo-gas-spiral}), the  stellar discs respond to the dark halo asymmetry via the global spiral pattern formation,  even if triaxiality in the halo is rather small ($\varepsilon = 0.02-0.05$).
 \item[3.]  The resulting amplitude of the spiral waves ($m=2$) depends on the triaxiality $\varepsilon$ and is practically independent of the initial disc conditions.

 \item[4.] Morphology of the spiral structure in our models is rather similar to that formed in classical density wave theory.
The growing spiral pattern does not lead to significant redistribution of mass in the disc in the
radial and azimuthal directions in the case of a non-axisymmetric massive halo.

 \item[5.] Triaxial halo does not result to significant dynamic heating of the stellar disc~(except, perhaps, the very initial stages of the interaction),
in contrast to the case of  a spiral pattern formed due to gravitational instability mechanisms.

\item[6.] Dynamics of the spiral pattern cannot be described by a single pattern angular velocity. Instead, we observe obviously two pattern angular velocities
at any fixed radius. Most of the time the spiral pattern at given radius rotates slowly with the angular velocity $ \Omega_{p1} $, but twice for it rotation cycle
its angular velocity jumps up to $\Omega_{p2} \simeq (4-5) \Omega_{p1}$ and then returns back.  These jumps is caused by passing
through the elongated potential well of the dark matter halo.

 \item[7.] In general, morphology and kinematics of the spiral pattern in a stellar disc are quite similar to those obtained in modelling of gaseous discs embedded in non-axisymmetrical dark matter haloes~\citep{Bekki-Freeman-2002!Spiral-Structure-Triaxial-Halo,Khoperskov-2012!Halo-gas-spiral}.

 \item [8.] There is a tendency for the halo to become more symmetric  in the disc plane in the models that take into account self-consistent interaction
 between the disc and the dark halo. In this case a formation of spiral density waves remains effective in the outer regions of disc even when the inner
 regions of the halo have become axisymmetric.
  \end{itemize}

\section{Acknowledgments}
The authors would like to thank the anonymous referee for helpful comments and advises that improved the paper. Authors express their gratitude for valuable discussions to V.I.~Korchagin. This work was supported by grants RFBR (11-02-12247, 12-02-31452, 12-02-00685-a, 13-02-01251) and the Program for State Support for Leading Scientific Schools of the Russian Federation (grant NSh-3602.2012.2).  This work was supported by the Federal Targeted Program of the Ministry of Education and Science of the Russian Federation "Research and Scientific-Pedagogical Personnel of Innovational Russia" for years 2009-2013.  Numerical simulations were run on supercomputers of NIVC MSU ``Lomonosov'' and ``Chebyshev''. S.~Khoperskov expresses his gratitude to non-commercial foundation ``Dynasty'' for financial support.

\end{document}